\documentclass[reprint, showpacs, amsmath, amssymb, aps, prb, superscriptaddress, floatfix]{revtex4-1}
\pdfoutput=1
\usepackage{xcolor}
\usepackage{graphicx}
\usepackage{dcolumn}
\usepackage{bm}
\usepackage{hyperref}
\usepackage{siunitx}
\usepackage{todonotes}
\usepackage{cleveref}
\usepackage{amsmath,amssymb}

\begin{document}
	\raggedbottom

\title{Growth of nematic susceptibility in the field-induced normal state of an iron-based superconductor revealed by elastoresistivity measurements in a 65~T pulsed magnet}

\author{J. A. W. Straquadine}
\thanks{Authors contributed equally}
\affiliation{Geballe Laboratory for Advanced Materials, Stanford University, California 94305, USA}
\affiliation{Department of Applied Physics, Stanford University, California 94305, USA}
\affiliation{Stanford Institute for Materials and Energy Sciences, SLAC National Accelerator Laboratory, 2575 Sand Hill Road, Menlo Park, CA 94025, USA}

\author{J. C. Palmstrom}
\thanks{Authors contributed equally}
\affiliation{Geballe Laboratory for Advanced Materials, Stanford University, California 94305, USA}
\affiliation{Department of Applied Physics, Stanford University, California 94305, USA}
\affiliation{Stanford Institute for Materials and Energy Sciences, SLAC National Accelerator Laboratory, 2575 Sand Hill Road, Menlo Park, CA 94025, USA}

\author{P. Walmsley}
\affiliation{Geballe Laboratory for Advanced Materials, Stanford University, California 94305, USA}
\affiliation{Department of Applied Physics, Stanford University, California 94305, USA}
\affiliation{Stanford Institute for Materials and Energy Sciences, SLAC National Accelerator Laboratory, 2575 Sand Hill Road, Menlo Park, CA 94025, USA}

\author{A. T. Hristov}
\affiliation{Geballe Laboratory for Advanced Materials, Stanford University, California 94305, USA}
\affiliation{Department of Physics, Stanford University, California 94305, USA}
\affiliation{Stanford Institute for Materials and Energy Sciences, SLAC National Accelerator Laboratory, 2575 Sand Hill Road, Menlo Park, CA 94025, USA}

\author{F.~Weickert}
\affiliation{National High Magnetic Field Laboratory (NHMFL), Los Alamos National Laboratory (LANL), Los Alamos, New Mexico 87545, USA}

\author{F. F. Balakirev}
\affiliation{National High Magnetic Field Laboratory (NHMFL), Los Alamos National Laboratory (LANL), Los Alamos, New Mexico 87545, USA}

\author{M. Jaime}
\affiliation{National High Magnetic Field Laboratory (NHMFL), Los Alamos National Laboratory (LANL), Los Alamos, New Mexico 87545, USA}

\author{R. McDonald}
\affiliation{National High Magnetic Field Laboratory (NHMFL), Los Alamos National Laboratory (LANL), Los Alamos, New Mexico 87545, USA}

\author{I. R. Fisher}
\affiliation{Geballe Laboratory for Advanced Materials, Stanford University, California 94305, USA}
\affiliation{Department of Applied Physics, Stanford University, California 94305, USA}
\affiliation{Stanford Institute for Materials and Energy Sciences, SLAC National Accelerator Laboratory, 2575 Sand Hill Road, Menlo Park, CA 94025, USA}

\date{\today}

\begin{abstract}	
	In the iron-based superconductors, both nematic and magnetic fluctuations are expected to enhance superconductivity and may originate from a quantum critical point hidden beneath the superconducting dome.
	The behavior of the non-superconducting state can be an important piece of the puzzle, motivating in this paper the use of high magnetic fields to suppress superconductivity and measure the nematic susceptibility of the normal state at low temperatures.
	We describe experimental advances which make it possible to measure a resistive gauge factor (which is a proxy for the nematic susceptibility) in the field-induced normal state in a 65~T pulsed magnet, and report measurements of the gauge factor of a micromachined single crystal of Ba(Fe$_{0.926}$Co$_{0.074}$)$_2$As$_2$ at temperatures down to 1.2~K.
	The nematic susceptibility increases monotonically in the field-induced normal state as the temperature decreases, consistent with the presence of a quantum critical point nearby in composition.
	
\end{abstract}

\maketitle
\section{Introduction}

	High magnetic fields suppress superconductivity, which makes it possible to study the low temperature properties of the less-understood electronic normal states from which unconventional superconductivity emerges.
	An important aspect of the normal state of several families of iron-based superconductors is the tetragonal-orthorhombic phase transition\cite{Hosono2015}, which originates from electronic correlations rather than a simple lattice instability.
	Tuning parameters such as chemical substitution or hydrostatic pressure suppress the critical temperature of the coupled nematic/structural phase transition, $T_S$, as well as the neighboring SDW transition temperature, $T_N$ and tends to increase the superconducting transition temperature $T_c$.
	It is currently unknown to what extent nematic and SDW fluctuations contribute to the low-temperature physics of Fe-based materials, nor even whether, for cases where the transitions are separated, the transitions remain continuous down to zero temperature.
	In the ``122'' family, such as Ba(Fe$_{1-x}$Co$_x$)$_2$As$_2$, the maximum $T_c$ occurs at approximately the same value of the tuning parameter at which $T_S$ and $T_N$ vanish, which suggests that superconductivity may be enhanced\cite{Labat2017} or even driven\cite{Lederer2017} by quantum critical fluctuations in the neighborhood of a quantum critical point (QCP) hidden beneath the superconducting dome.

	Numerous experiments, including quantum oscillations\cite{Walmsley2013}, NMR\cite{Ning2010}, London penetration depth\cite{Hashimoto2012}, shear modulus\cite{Yoshizawa2012}, resistivity\cite{Doiron-Leyraud2009, Chu2010, Licciardello2019}, and elastoresistivity \cite{Chu2012, Kuo2016} point towards the presence of a QCP in several ``122'' compounds, although a definitive connection to superconductivity has yet to be established.
	Measurements of the nematic susceptibility to very low temperatures for compositions proximate to the putative quantum critical point may shed some light on the relevant fluctuations.
	In the present work we focus specifically on the nematic susceptibility, demonstrating new  means to measure this quantity for a nearly optimally doped composition of Ba(Fe$_{1-x}$Co$_x$)$_2$As$_2$ while superconductivity is suppressed by a large magnetic field.
	While estimates of the nematic susceptibility have been obtained previously to relatively low temperatures for similar compositions via elastic stiffness\cite{Bohmer2014} and Raman\cite{Wu2017b} measurements, those measurements were performed in zero magnetic field within the superconducting state.
	Here, we study the temperature dependence of the nematic susceptibility  in the absence of superconductivity down to a temperature of 1.2~K via an elastoresistivity technique.

	Elastoresistivity is a fourth-rank linear response tensor characterizing the sensitivity of the resistivity to strain.
	Specific components of the elastoresistivity tensor are directly proportional to the nematic susceptibility\cite{Shapiro2016a}, and such measurements can therefore detect effects of nematic fluctuations in the tetragonal state
	Due to its nature as an electrical resistance measurement, however, elastoresistivity cannot be measured in the superconducting state and any low temperature studies must rely on an external mechanism such as magnetic field, or chemical disorder to suppress $T_c$.
	An added challenge of working with Co-doped BaFe$_2$As$_2$ near optimal doping is that the high upper critical field at zero temperature $H_{c2} (0)=52$~T is beyond what is currently accessible with static magnetic fields.
	A measurement of the elastoresistivity in the field-induced normal state near a putative QCP must therefore be done in a pulsed magnet.	
	This sets a stringent constraint on the timescales of the measurement and requires a new experimental approach.
	
	The purpose of this work is twofold.
	First, we present adaptations to existing elastoresistivity techniques\cite{Hristov2018} which extend the range of applicability of elastoresistivity measurements to higher magnetic fields and lower temperatures.
	Second, we apply this technique for the first time to a sample of near-optimally doped Ba(Fe$_{1-x}$Co$_x$)$_2$As$_2$, where $x=0.074$.
	We show that the resistive gauge factor, which acts as a proxy for the nematic susceptibility, increases monotonically as temperature decreases down to our base temperature of 1.2~K.
	As we will also show, we observe no significant field dependence of the elastoresistivity, indicating that the driving force behind the nematic fluctuations is not strongly altered by extreme magnetic field.
	This work provides a new experimental perspective on the important region of the phase diagram close to optimal doping.
	At a minimum, for this composition close to optimal doping, when the superconductivity is suppressed by an external magnetic field, the continued growth of the nematic susceptibility as temperature is reduced towards absolute zero is not inconsistent with the presence of a QCP with a nematic character nearby in composition.
	Combined with other insights, this observation adds to the body of evidence that suggests that nematic fluctuations might play an important role in the low temperature physics at or near optimal doping in the iron pnictides.

\section{Experimental Methods}
    \begin{figure*}
        \centering
        \includegraphics[width=\textwidth]{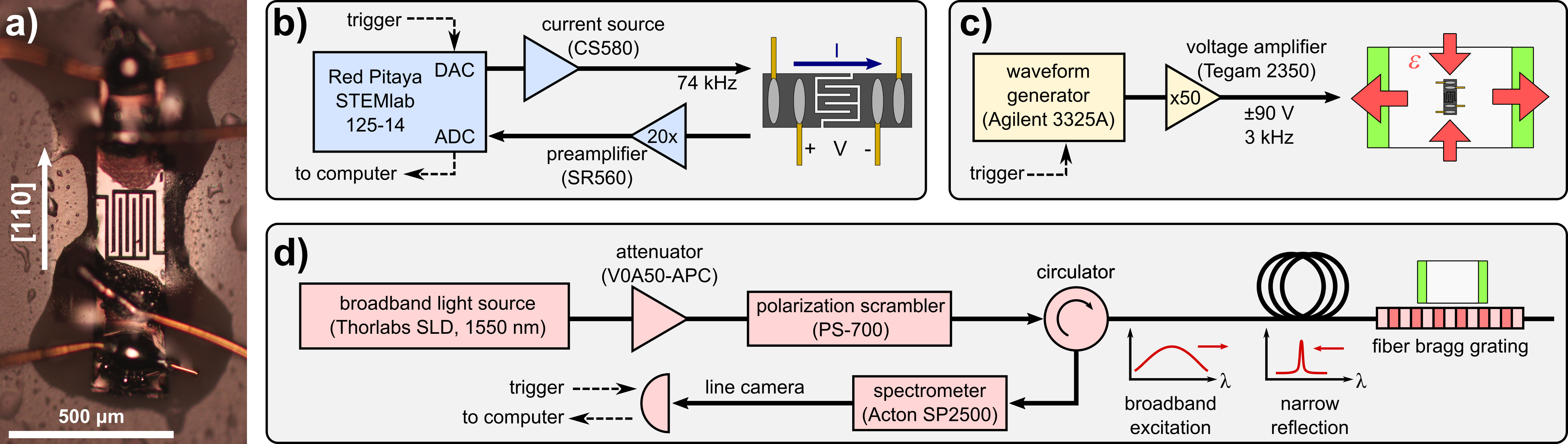}
        \caption{Schematic diagram of the experiment. 
        	Panel a) shows an optical microscope image of the sample after FIB micromachining. The sample is adhered to the surface of the PZT stack, and the four contacts on the top and bottom of the image are embedded in epoxy.  
        	Panel b) shows the setup for the resistance measurement.  The excitation signal is generated by the Red Pitaya STEMlab~125-14 device, and the sample voltage is detected by the same device at a digitization rate of 125~MHz.  
        	Panel c) shows the circuit used to drive the piezoelectric device.  The waveform generator is synchronized to produce 18~cycles of a sine wave starting 6 ms after the beginning of the pulse.  
        	Panel d) shows the optical path for the fiber Bragg grating (FBG) strain sensor.  A continuous wave, broad-spectrum LED light source illuminates the grating, and the grating only reflects in a narrow band of wavelengths.  A spectrometer and line camera records the spectra during the pulse at a frequency of 46.5~kHz.
    	}
        \label{fig:schematic}
    \end{figure*}

    The sample of Ba(Fe$_{0.926}$Co$_{0.074}$)$_2$As$_2$ was grown using a self-flux technique described in detail elsewhere\cite{Chu2009}. The Co doping concentration was measured using electron probe microanalysis (EPMA) in a JEOL JXA-8230 SuperProbe system calibrated using the parent compound, BaFe$_2$As$_2$, and Co standards.  The sample was cleaved and cut into a rectangular bar with edges along the tetragonal [110] and [1$\bar{1}$0] crystallographic axes with dimensions \SI{430}{\micro\meter}$\times$\SI{1950}{\micro\meter}$\times$\SI{13}{\micro\meter}.
    Four electrical contacts were made to the sample with \SI{25}{\micro\meter} diameter gold wire and Chipquik SMD291AX10T5 Sn63/Pb37 solder beads.  The full contacting process is described in depth elsewhere\cite{Ikeda2018}.
    The contacted sample was glued using an AngstromBond epoxy (AB9110LV) onto a piezoelectric stack (Piezomechanik PSt150/5x5/7 cryo 1) such that the tetragonal $ab$ plane was flush with the face of the piezoelectric stack and the long axis of the sample was perpendicular to the stack's poling axis.
    Care was taken to use only a small amount of glue so the sample's top surface remained clean.  The glue was cured by baking at $45^\circ$C for 5-6 hours.
    
    Throughout this paper we work in a coordinate system aligned with the PZT stack: with the $x$ axis defined along the sample, $y$ axis along the PZT stack poling axis, and $z$ parallel to the out-of-plane crystallographic $c$ axis.
    This is described graphically in the inset to \cref{fig:results}.
    In this coordinate system, which is rotated by 45$^\circ$ about $z$ with respect to the in-plane crystallographic axes, tensor quantities with $x^2-y^2$ symmetry belong to the $B_{2g}$ irreducible representation of the $D_{4h}$ crystallographic point group.
    
    In order to increase the size of the signal and increase the signal-to-noise ratio, we increased the resistance of the sample using focused ion beam (FIB) micromilling.  After the sample has been contacted and adhered to the PZT stack, the sample was patterned into a long meander.  Milling was performed using an FEI Helios NanoLab 600i DualBeam FIB/SEM.  An optical microscope image of the resulting pattern is shown in \cref{fig:schematic}a.  Each of the seven bars is \SI{150}{\micro\meter} long, \SI{12.5}{\micro\meter} wide, and the entire sample is \SI{13}{\micro\meter} thick.  Machining the sample in this way increases its resistance by a factor of approximately 90, to a final value of 25~$\Omega$ at room temperature.
    After machining, the cuts in the sample are filled in with the same low-viscosity epoxy used to adhere the sample to the stack.  The stack was then loaded into a 65~T multi-shot magnet at the National High Magnetic Field Lab Pulsed Field Facility at the Los Alamos National Laboratory.  The sample and PZT stack are mounted such that the magnetic field is applied perpendicular to the $ab$ plane of the sample.
    
    A significant source of signal interference in pulsed magnets is the voltage induced in the sample and wires by the rapidly changing magnetic field.  In order to minimize the impact of this effect, we mounted the PZT stack and sample onto a homebuilt stage consisting of an electromagnet coil which was driven with an oscillating current.  We then measured the magnitude of the inductive pickup while manipulating the wires, and shaped the wires to minimize this signal. The inductive pickup effect is almost completely canceled, except for a sharp spike at the beginning of the pulse which is suppressed by the phase-sensitive detection and filtering.
    
    \Cref{fig:schematic}b shows a schematic of the experimental setup for measuring the sample resistance. Both the voltage measurement and sample excitation were accomplished using a Red Pitaya STEMlab~125-14 system, which is a multipurpose data acquisition board based on the Xilinx Zynq-7000 family of FPGAs.  The excitation signal for the sample was generated by the digital-to-analog converter (DAC) converted to a current using a Stanford Research Systems CS580 voltage to current converter.  The sample voltage was amplified by a Stanford Research SR560 preamplifier with a gain of 20~V/V, then detected by the STEMlab's 14-bit ADC at a sampling rate of 125~MHz, stored in memory on the board, and then transferred to another computer after the pulse.
    
    \begin{figure}
        \centering
        \includegraphics[width=\columnwidth]{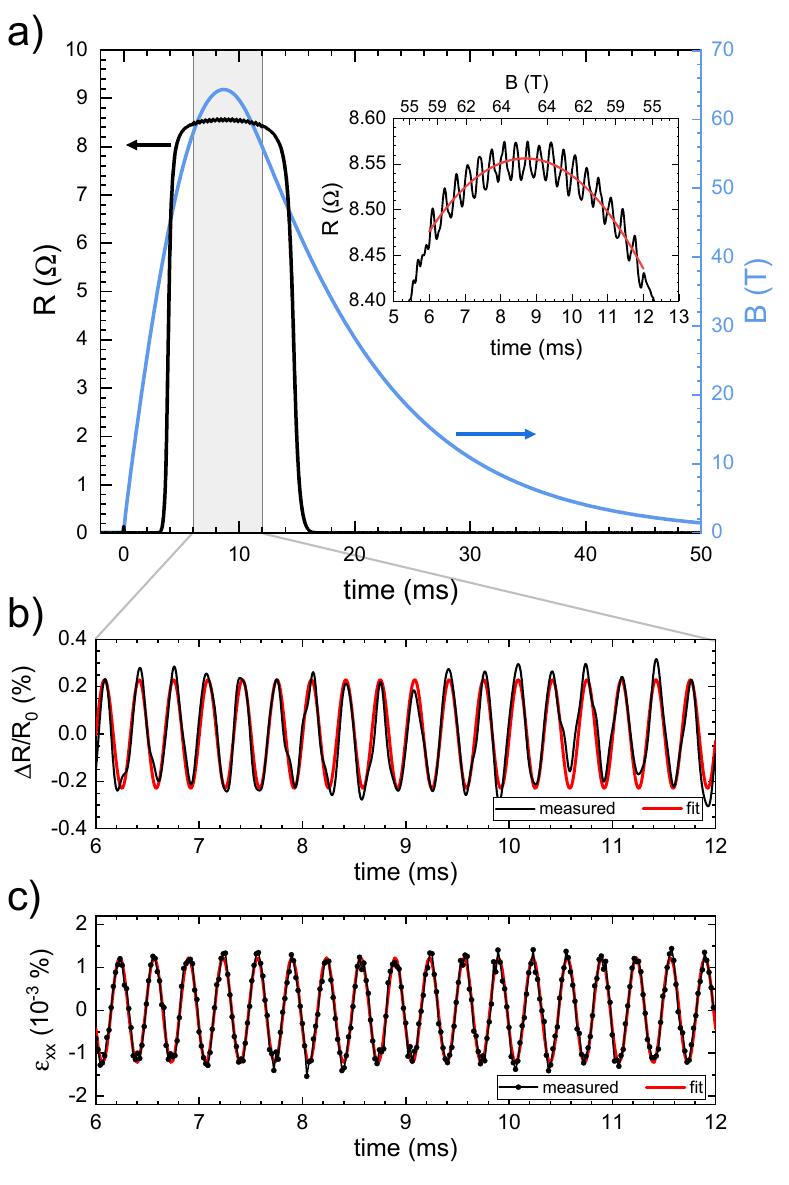}
        \caption{A representative elastoresistivity measurement of Ba(Fe$_{0.926}$Co$_{0.074}$)$_{2}$As$_{2}$ during a magnetic field pulse.
        This trace was taken at 3~K, well below the zero-field superconducting $T_c$, and superconductivity was fully suppressed with a 65~T pulse.
        (a) The field (blue line) and amplitude of the sample response (black line) during the magnetic field pulse.
        The magnet is fired at the t=0~ms.
        When the field surpasses $H_{c2}\approx 50T$, the sample becomes resistive.
        The PZT stack is energized with a 3kHz drive for 6~ms around peak field (gray shaded region).
        Inset: magnification of the sample response at peak field.
        (b) The sample response (black line) after subtraction of the magnetoresistance background while the PZT stack is driven.
        The resistance of the sample oscillates in time due to the elastoresistivity response.
        This is fit with a 3~kHz sine wave (red line).
        (c) The strain (black line) measured by the FBG grating while the PZT stack is driven.
        This is also fit by a 3~kHz sine wave (red line).}
        \label{fig:analysis}
    \end{figure}
    
	An elastoresistivity measurement requires that the sample resistance and the strain be detected simultaneously.  We measure the strain through optical spectroscopy \cite{Jaime2017} of a fiber Bragg grating (FBG).  We use a grating with peak wavelengths in the range of 1550~nm.  The fiber is illuminated with polarization-scrambled light from a wide-spectrum LED, and the reflected light is analyzed using a Princeton Instruments Acton SP2500 spectrometer and line camera.  The strain is extracted from the shift in the peak position in the spectrum of the reflected light, using a calibration factor of 0.0012~nm/ppm strain.  The fiber contains several gratings spaced by several millimeters; we adhere one to the PZT and leave the rest freestanding.  These freestanding gratings are not affected by the PZT strain but are otherwise subject to the same environment, so we subtract the peak shift from these gratings to eliminate magneto-optical effects, thermal drifts, mechanical vibrations, or other spurious errors.  The FBG is adhered to the side of the PZT stack parallel to the field and $c$-axis of the sample using Stycast 2850FT blue epoxy.  With this configuration, both the long axis of the sample and FBG are aligned orthogonal to the poling axis of the PZT, and the strain measured by the FBG can be used as a proxy for the strain experienced by the sample.  Further checks with resistive strain gauges show that strain in the two axes is indeed the same to within 10\% at all temperatures, as shown in \cref{app:strainxz}.
	
	The extraction of the gauge factor during the pulse is achieved using an oscillating strain, similar to the technique described in ref. \onlinecite{Hristov2018}.  We apply a sinusoidal excitation current of 2~mA\textsubscript{rms} at 74.4048~kHz into the sample, and we drive the PZT with a sinusoidal voltage of $\pm$90~V at 3~kHz.  The drive voltage begins oscillating 6~ms after the beginning of the magnetic pulse, and continues for 6~ms, or 18~cycles, such that the PZT is only driven in a small window centered on the peak magnetic field.  This prevents heat generated by dissipation in the PZT stack from raising the temperature of the sample.
	Based on measurements of the critical field $H_{c2}$ before and after the pulse, as seen in \cref{app:heating}, the heat generate by the PZT does not significantly heat the sample on the timescale of a pulse.  A representative measurement can be seen in \cref{fig:analysis}a.
	
	Once the signal is acquired, we then use a software lock-in amplifier to perform the amplitude demodulation and extract the changes due to strain.  The digital lock-in consists of multiplying the raw signal with a synthesized sine wave of unit amplitude at the sample current frequency, followed by a low-pass filter.  In this analysis we use the built-in MATLAB infinite impulse response low-pass filter function, with a cutoff frequency of 4321~Hz and a roll-off of 18.54~dB/decade.  There is, however, a broad range of appropriate filter parameters.
	
	Aside from the change in resistance due to the oscillating strain, the sample resistance also changes due to the magnetoresistance of the sample and roughly follows the shape of the magnetic field pulse itself.
	This effect is largest at the lowest temperatures and largest fields where the zero strain resistance changes by $\pm$2\% around the average resistance during the 6~ms window while the PZT stack is driven.
	The magnetoresistance appears roughly linear in this small window near peak field, but is not inconsistent with quadratic magnetoresistance near 0~T as observed at higher temperatures\cite{Rullier-Albenque2013}.
	We compensate for this effect by subtracting a quadratic background from the resistance signal (inset of \cref{fig:analysis}a).
	The strain-induced change in resistance as well as the strain itself extracted from the FBG (\cref{fig:analysis}c) are then each fit by a sine wave at the strain frequency.  The fitted amplitudes, $\Delta R$ and $\varepsilon_{xx}$ respectively, and the average zero applied stress resistance $R_0$\footnote{In this analysis we have chosen to normalize the change in resistivity due to strain by the resistivity of the sample in the normal state.  This is the average measured resistivity of the sample during a pulse and requires no extrapolation.  If the physical origin of the magnetoresistance of a sample was independent of the nematic fluctuations an argument could be made to normalize by the extrapolated zero field resistance.  A comparison is shown in \cref{app:norm}.  For the family of materials discussed here the magnetoresistance is small and this is only a small quantitative correction.  The choice of normalization does not affect any of the conclusions drawn here.}  are used to calculate the gauge factor, $G = (-\Delta R/R_0)/\varepsilon_{xx}$.
	We incorporate a negative sign such that $G$ is always positive for our sample: for Ba(Fe$_{1-x}$Co$_x$)$_2$As$_2$ the resistivity decreases under tensile (positive) strain\cite{Chu2010}.
	
	The sample temperature is monitored before and after the pulse by a cernox temperature sensor, and the upper critical field $H_{c2}$ of the sample itself is also used as a secondary local thermometer.  The method for extracting $H_{c2}$ and converting it to temperature is described in \cref{app:heating}, and the reported temperatures throughout this work incorporate this correction.

\section{Results and Discussion}

    \begin{figure}
        \centering
        \includegraphics[width=\columnwidth]{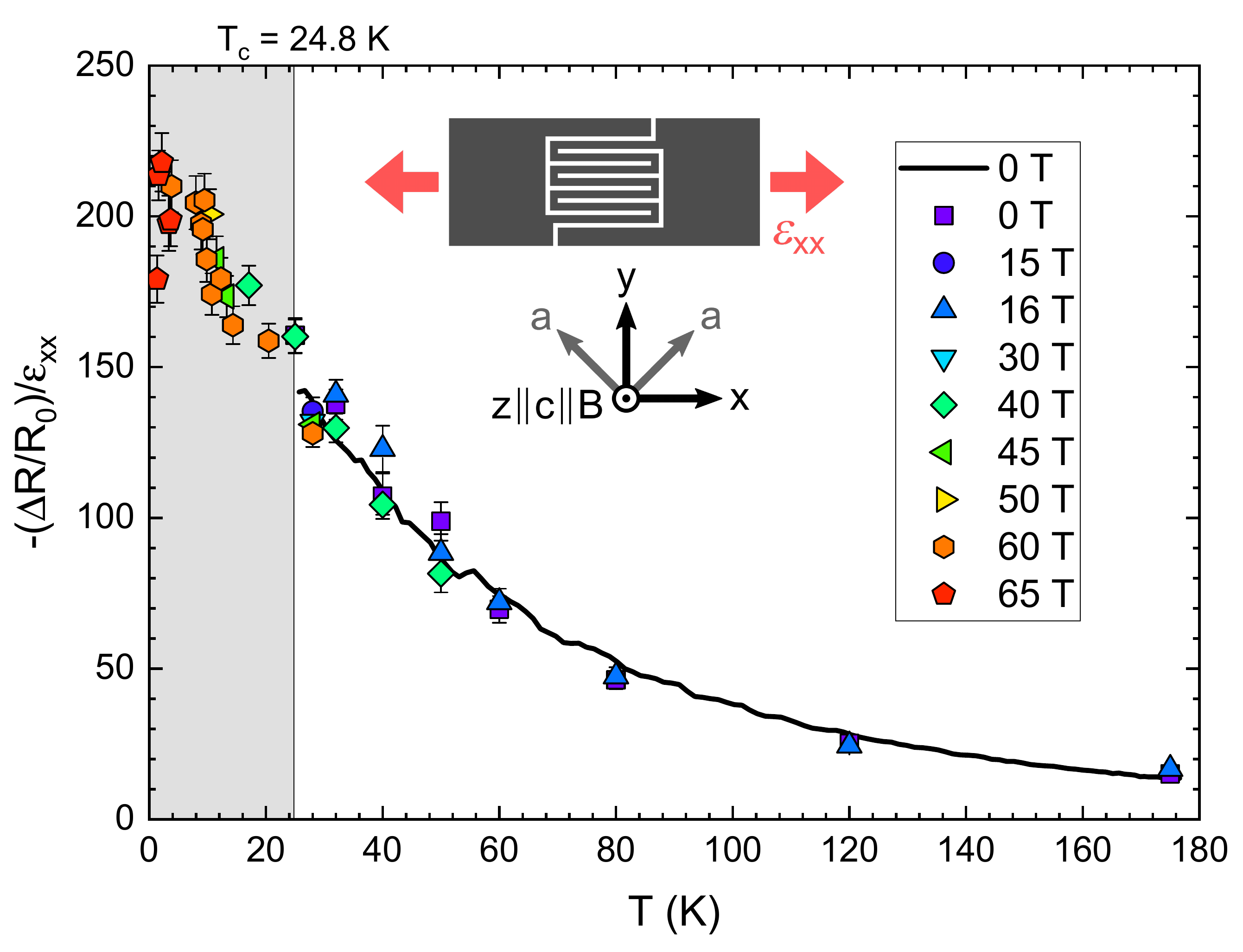}
        \caption{The extracted resistive gauge factor of a micromachined sample of Ba(Fe$_{0.926}$Co$_{0.074}$)$_2$As$_2$ as a function of temperature and field.  Data taken during a pulse is represented by filled symbols and zero-field data taken using the continuous AC elastoresistivity technique \cite{Hristov2018} is represented by a solid black line.  Error bars represent one standard deviation, as calculated in detail in \cref{app:errors}  Measurements at all field values produce the same result within error bars, and match with the zero-field data at temperatures above $T_c$=24.8~K.  Below the zero field superconducting transition (vertical black bar) the gauge factor (and therefore the nematic susceptibility) continues to increase smoothly with decreasing temperature, indicating an increasing importance of nematic fluctuations.
        Note that we  use a rotated coordinate basis (inset) in which the cartesian $x$ and $y$ axes are roated by 45$^\circ$ with respect to the in-plane crystal axes.}
        \label{fig:results}
\end{figure}

The measured gauge factor $G=(-\Delta R/R_0)/\varepsilon_{xx}$ as a function of temperature is shown in \cref{fig:results}.
Each filled symbol represents a single pulse of the magnet.
As the temperature decreases, the response of the sample resistance to strain increases in magnitude.
For temperatures above the zero-field superconducting transition $T_c$=24.8~K, we also measured the gauge factor at zero field for comparison.
This trace (black line in \cref{fig:results}) was taken using 0.251~Hz strain.
Our extraction of the gauge factor from the pulsed measurements agrees well with the conventional technique, despite the four orders of magnitude difference in frequency.
This is consistent with previous measurements in 2.5\% Co-doped Ba-122 \cite{Hristov2018}.

The pulsed magnetic field and the cryostat make it possible to measure the gauge factor down to 1.2~K, which is more than an order of magnitude closer in temperature to the putative quantum critical point than previous measurements of the elastoresistivity in the iron-based superconductors.
At lower temperatures, when superconductivity has been suppressed by magnetic field, we find that the gauge factor continues to increase smoothly and monotonically.
\footnote{At the coldest temperature, 1.2~K, the elastoresistivity decreases slightly.  It is $4\sigma$ below the nearest data point (1.6~K).  This is at the very limit of our accessible temperature range and we can just barely suppress superconductivity in the full 65T field.  Due to these limitations we cannot draw any conclusions from this apparent decrease without further measurements.}
This is our main result.

Strictly speaking, the resistive gauge factor reported here is related to, but not exactly equal to, the $B_{2g}$ nematic susceptibility $\chi_N$.  An ideal measurement of elastoresistivity components proportional to $\chi_N$ would measure the two in-plane components of the resistivity tensor $\rho_{xx}$ and $\rho_{yy}$ independently, as well as the strain components $\varepsilon_{xx}$ and $\varepsilon_{yy}$. Altogether, this would then permit decomposition of the measured elastoresistivity into its various independent symmetry channels, and in particular,
\begin{equation}
	m_{B_{2g}} = m_{xxxx}-m_{xxyy} =  \frac{\Delta\rho_{xx}-\Delta\rho_{yy}}{\rho_0 (\varepsilon_{xx}-\varepsilon_{yy})} \propto \chi_N. 
\end{equation}
In this work, the single meandering resistance bar used to maximize the signal is primarily sensitive to $\rho_{xx}$; finite element modeling of current flow through the sample geometry show that the measured resistance comprises 93\%~$\rho_{xx}$, and 7\%~$\rho_{yy}$.  A single resistance measurement does not allow for independent determination of $\rho_{yy}$.  However, based on prior measurements, we can make several general statements which justify the use of the measured gauge factor as a proxy for the nematic susceptibility.

To linear order, the x-axis resistivity $\rho_{xx}$ in a system with tetragonal symmetry can be affected by antisymmetric strain $\varepsilon_{B_{2g}}=(\varepsilon_{xx}-\varepsilon_{yy})/2$ and two different forms of symmetry-preserving strains:
isotropic in-plane strain $\varepsilon_{A_{1g,1}}=(\varepsilon_{xx}+\varepsilon_{yy})/2$  and out-of-plane strain $\varepsilon_{A_{1g,2}}=\varepsilon_{zz}$.
Strains with a $B_{1g}$ symmetry (i.e. antisymmetric strain rotated by 45$^\circ$ about the $c$ axis) cannot contribute to linear order due to the orientation of the sample.
The gauge factor $G$ contains contributions from the elastoresistivity coefficients of various symmetries weighted by in-plane Poisson ratio $\nu=\varepsilon_{yy}/\varepsilon_{xx}$ and out-of-plane Poisson ratio $\nu_z=\varepsilon_{zz}/\varepsilon_{xx}$ according to the relation
\begin{equation}\label{eq:g}
	G = \left(\frac{1-\nu}{2}\right)m_{B_{2g}} + \left(\frac{1+\nu}{2}\right)m_{A_{1g,1}} + \left(\frac{\nu_z}{2}\right)m_{A_{1g,2}}.
\end{equation}
where $m_{A_{1g,1}} = \frac{\Delta\rho_{xx}+\Delta\rho_{yy}}{\rho_0 (\varepsilon_{xx}+\varepsilon_{yy})}$ and $m_{A_{1g,2}} = \frac{\Delta\rho_{xx}+\Delta\rho_{yy}}{\rho_0 (\varepsilon_{zz})}$.
In any strain experiment, the Poisson ratio which should enter into \cref{eq:g} depends not only on the elasticity tensor of the material, but also on the boundary conditions imposed on the sample.  Some care is required in understanding the proper effective Poisson ratio.
In the limit of a thin film sample, the appropriate value of $\nu$ is given by that of the PZT stack, which has been measured to be $\nu \approx -2.3$ at low temperatures, although it is temperature dependent \cite{Hristov2018}.
The opposite limit, in which the sample is treated as a free-standing beam and compressive or tensile stress is applied to both ends, is controlled by the Poisson ratio of the material itself, which is $\nu\approx-0.26$.\cite{Ikeda2018}
Due to the complex situation of a micromachined sample embedded in epoxy, we expect the Poisson ratio to lie between these two limits, although we do not have a direct measure.
\footnote{The massive apparent difference between the magnitudes of the Poisson ratios discussed here stems from a difference of which strain axis ought to be considered ``primary.''  In our experimental setup, the poling axis of the PZT stack runs along $y$, producing $|\varepsilon_{yy}|>|\varepsilon_{xx}|$.  On the other hand, if the encapsulating epoxy is not stiff enough to transmit this strain to the sample (which has, within the meandering structure, a cross-sectional aspect ratio of approximately unity) then the primary effect would be a uniaxial stress along the $x$ axis (the long axis of the sample) producing $|\varepsilon_{xx}|>|\varepsilon_{yy}|$.  The crossover between these limits corresponds to $\nu=-1$.}

Despite the uncertainty in the exact value of $\nu$ that characterizes the strain in the meander, it can be shown that the antisymmetric strain component $\varepsilon_{B_{2g}}$ is larger by a factor of at least 1.7 for all possible values of $\nu$, and that the $A_{1g}$ coefficient $(1+\nu)/2$ in fact vanishes at a crossover value of $\nu=-1$.
Moreover, prior measurements\cite{Palmstrom2017} have demonstrated that the response of the in-plane resistivity of Ba(Fe$_{1-x}$Co$_x$)$_2$As$_2$ to in-plane isotropic in-plane strains ($m_{A_{1g,1}}$) as well as out-of-plane strains ($m_{A_{1g,2}}$) is smaller than $m_{B_{2g}}$ by an order of magnitude in underdoped Ba(Fe$_{1-x}$Co$_x$)$_2$As$_2$ and does not have a strong temperature dependence.
Considering all of these factors, the response to antisymmetric strain, $m_{B_{2g}}$, which has been shown previously to correspond to the nematic susceptibility\cite{Chu2012, Riggs2014, Shapiro2016a} is expected to provide the dominant contribution to the gauge factor.

Two additional factors limit the quantitative accuracy of the present technique to estimate $m_{B_{2g}}$.
First, the strain within the meandering section of the sample, which is embedded within epoxy, may be both inhomogeneous and may also deviate from the strain in the PZT stack.  Several studies have examined the transmission of strain from PZT stacks through layers of epoxy and into pnictide samples \cite{Kuo2016, Palmstrom2017}, but these focused on large, flat samples adhered with different epoxies than the one used in this work.  While the absolute magnitude of the local strains is unknown, the \emph{change} of the strain environment when the PZT stack is energized will still carry primarily $B_{2g}$ character as a consequence of the relative alignment of the stack and the sample.

Secondly, PZT stacks exhibit strongly anisotropic thermal expansion in which the stack expands along the $y$ axis upon cooling while contracting along the $x$ axis.
The resulting nonzero offset in $\varepsilon_{B_{2g}}$ can result in measurable contributions from higher order elastoresistivity responses.
In particular, the response of the $A_{1g,1}$ component of the resistivity tensor to the square of the antisymmetric strain $\varepsilon_{B_{2g}}^2$ has been shown to diverge on approach to the nematic transition in 2.5\% Co-doped BaFe$_2$As$_2$.\cite{Palmstrom2017}
A nonzero $B_{2g}$ offset strain can therefore affects the apparent linear response of the isotropic resistivity to a small perturbation and alters the measured gauge factor.
However, in compositions where the quadratic coefficient has been measured, the quadratic effect has been shown to depend even more strongly on temperature than the linear part, such that any effects present in this measurement should alter the observed temperature dependence.
This is not observed in the present measurement.
Further measurements of the quadratic coefficients near optimal doping range must be made in order to clarify the effect of offset strains.

These complexities notwithstanding, we find that the temperature dependence of our data compares very well with prior work\cite{Chu2012, Kuo2016} in zero field above $T_c$.
A direct comparison is shown in \cref{app:historical}.
In short, the gauge factor here exhibits the same temperature dependence as the $B_{2g}$ elastoresistivity component and hence the nematic susceptibility.

Our data clearly reveal that the nematic susceptibility continues to grow in a smooth and continuous manner with decreasing temperature, without any observable cusp or saturation.
The role played by critical fluctuations in enhancing the nematic susceptibility can in principle be further understood by analyzing the precise functional form of the susceptibility at low temperatures.
Realistic parameters for a Curie-Weiss dependence of the susceptibility with temperature (as may be expected from a mean-field continuous phase transition\cite{Kuo2016}), however, are unable to fit our data over the full range of temperatures.
The lowest temperature data does not diverge strongly enough, as demonstrated in \cref{app:cw}; in particular, the temperature dependence of the gauge factor within the field-induced normal state appears approximately linear in temperature.
This behavior could perhaps be explained either by quenched disorder limiting the correlation length of quantum critical fluctuations or by Landau damping from metallic degrees of freedom \cite{Schattner2016, Lederer2017, Berg2019}.

It is worth noting that magnetic fields are known to have little effect on $T_S$\cite{Chu2010a}, an effect which is corroborated by the field independence of the gauge factor.
Deviations from a simple scaling due to field-induced motion of the transition itself are therefore unlikely.
Also, we note that magnetic fields up to 65~T do not appear to significantly perturb the nematic fluctuations in this material.

Also, the featureless nature of the susceptibility curve demonstrates that no other competing phases transitions occur as superconductivity is suppressed.  This suggests that while a 65~T magnetic field destroys superconductivity, the high temperature normal state and the field-induced normal state are indeed adiabatically connected.  The main result is that we infer that the nematic susceptibility continues to rise to the lowest temperatures once superconductivity is suppressed.  This is consistent with the presence of a QCP with a nematic character nearby in composition space.  The role played by the strong nematic fluctuations in terms of the superconductivity and also other properties remains open, but our observation, taken together with other probes of the nematic fluctuations in this system, is highly suggestive that these might be connected. 

\section{Conclusions}
We have presented a new implementation of elastoresistivity measurements in the iron-based superconductors to high fields (65~T) and low temperatures (1.2~K).
We then used this technique to extract a resistive gauge factor, a proxy for the nematic susceptibility, in the field-induced normal state at low temperatures of a micromachined sample of Ba(Fe$_{0.926}$Co$_{0.074}$)$_2$As$_2$.
The gauge factor grows smoothly and monotonically as the temperature decreases down to the lowest attainable temperatures, which supports the notion that strong nematic fluctuations in the non-superconducting state of the iron pnictides may stem from a quantum critical point hidden beneath the superconducting dome.
This work provides the first step in mapping the nematic susceptibility in the iron pnictide superconductors near the putative quantum critical point.

\section{Acknowledgements}
EPMA measurements were taken at the Stanford Microchemical Analysis Facility with the help of Dr. Dale Burns. This work was supported by the Department of Energy, Office of Basic Energy Sciences, under Contract No. DE-AC02-76SF00515. A portion of this work was performed at the National High Magnetic Field Laboratory, which is supported by the National Science Foundation Cooperative Agreement No. DMR-1644779 and the State of Florida.
Part of this work was performed at the Stanford Nano Shared Facilities (SNSF) supported by the National Science Foundation under award ECCS-1542152.
J.S. acknowledges support as an ABB Stanford Graduate Fellow.
J.C.P. was supported by a Gabilan Stanford Graduate Fellowship and a Stanford Lieberman Fellowship during the course of this work.
J.C.P. and A.T.H. were supported by a NSF Graduate Research Fellowship (grant DGE-114747).

\pagebreak
\bibliographystyle{apsrev4-1}
\bibliography{PulsedFieldER}

\appendix

\section{Normalization of resistivity ratios}\label{app:norm}
\begin{figure}
        \centering
        \includegraphics[width=8cm]{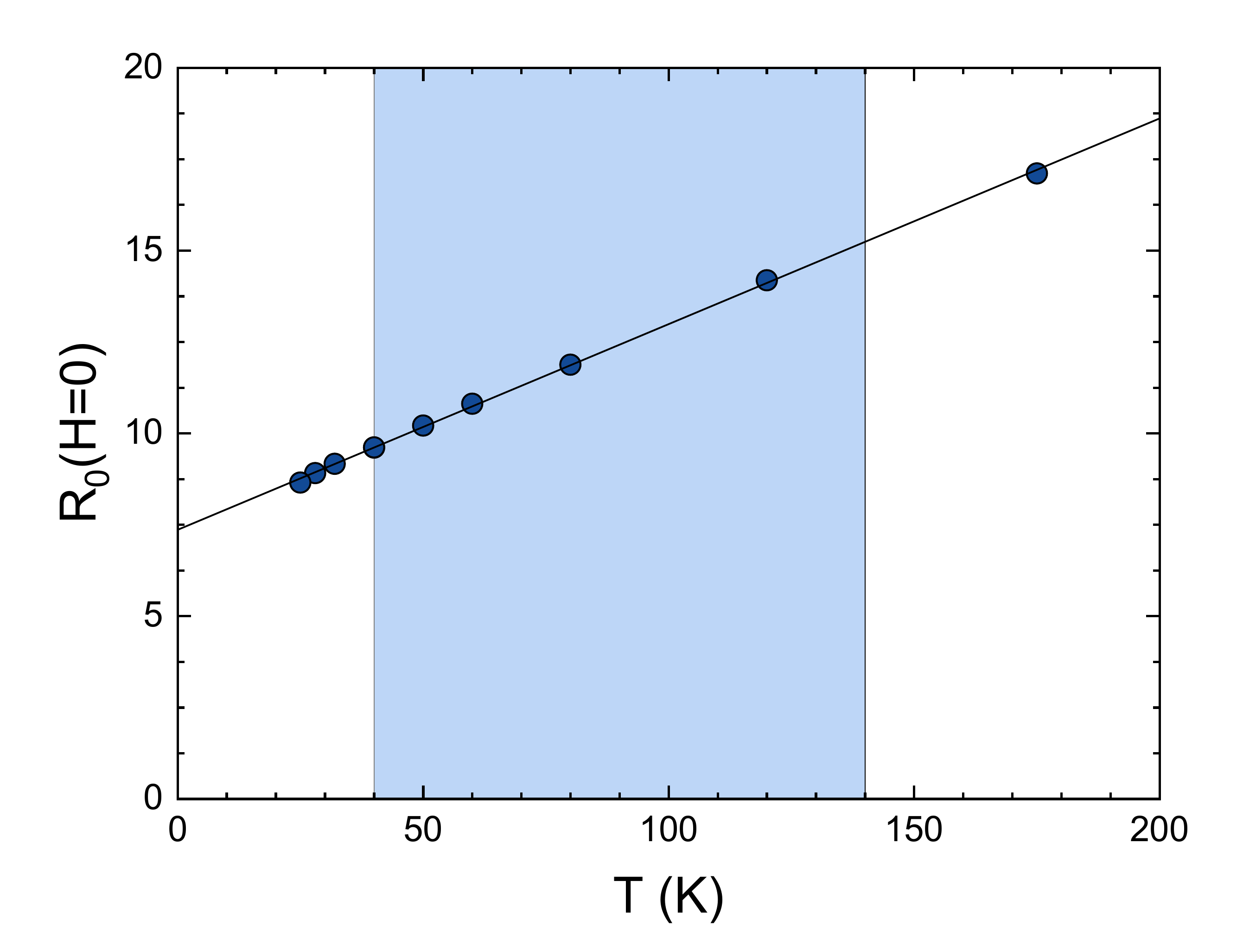}
        \caption{Linear extrapolation of the zero field resistivity down to cold temperatures.
        The deviation between the measured resistance at 65~T and the extrapolated value at the same temperature is largest at the coldest temperatures and highest fields, at 12\%.
    	The shaded region is the region over which the linear fit was performed.}
        \label{fig:res_extrap}
    \end{figure}

\begin{figure}
        \centering
        \includegraphics[width=8cm]{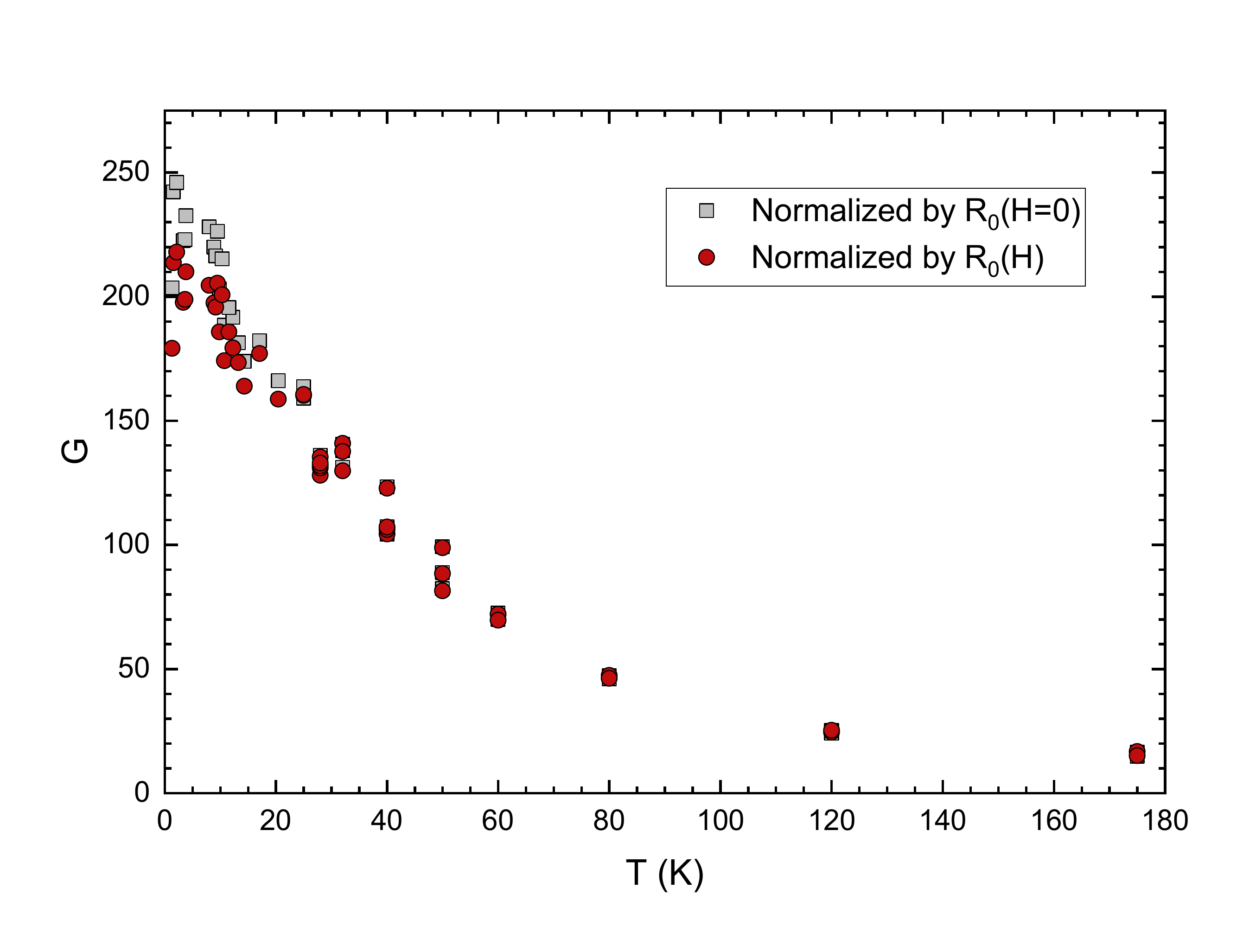}
        \caption{Comparison of the extracted gauge factor $G=(-\Delta R/R_0) /\varepsilon_{xx}$ using $R_0=R_0(H)$ and $R_0=R_0(H=0)$.
        	The two differ only qualitatively, and the choice of normalization does not change the conclusions we draw from the data.}
        \label{fig:res_NormComp}
    \end{figure}

	The definition of the base resistance $R_0$ used to normalize the elastoresistivity response carries some ambiguity.\cite{Shapiro2015}  If the physical origin of the magnetoresistance of Ba(Fe$_{1-x}$Co$_x$)$_2$As$_2$ were independent of nematic fluctuations, the most physically motivated choice would be to normalize by the extrapolated zero field resistance.  Near optimal doping the resistance as a function of temperature can be well fit by a straight line, as shown in \cref{fig:res_extrap}.  We can approximate $R(H=0)$ in the field-induced normal state by simply extrapolating this fit to lower temperatures.
	In our analysis, however, we calculate the gauge factor $(-\Delta R/R_0)/\varepsilon_{xx}$ using the average resistance measured at peak field $R_0(H)$.  This quantity is chosen because it can be directly measured during the magnetic field pulse and requires no extrapolation nor assumptions about the origin of the magnetoresistance.  In any case, the magnetoresistance of Ba(Fe$_{1-x}$Co$_x$)$_2$As$_2$ is small \cite{Rullier-Albenque2013}.  At the lowest temperatures and highest fields there is a quantitative difference (up to 12\% in the field and temperature ranges considered in this study) between $R_0(H)$ and the extrapolated low temperature $R_0(H=0)$.	A comparison of the normalizations is shown in \cref{fig:res_NormComp}.  The differences do not affect the conclusions we draw from this work. 
	
\section{Sample heating detected through $H_{c2}$}\label{app:heating}
    In our implementation of the elastoresistivity measurement, the sample is directly adhered to the surface of the PZT device.  At high operation frequencies, however, piezoelectric devices are known to generate significant heat which could affect our measurement.  In order to verify the temperature of the sample independent of thermometry errors, we extracted $H_{c2}$ from the resistivity curves.  This extraction can be done both on the increasing and decreasing field sweeps.  We observe a temperature offset at temperatures above the boiling point of liquid helium but below 10~K.  Also, there is a slight decrease in $H_{c2}$ in the downsweep relative to the upsweep, which indicates some slight heating of the sample during the pulse.  By performing similar measurements without driving the PZT stack and using only 10\% of the excitation current to the sample (1\% the excitation power in the resistive phase) we see that most of this heating still occurs, suggesting that it is not due to either the PZT or Joule heating within the sample or contacts.  Vortex pinning effects are unlikely to explain the change in $H_{c2}$ considering that the effect disappears once the sample is submerged in liquid helium.  We therefore attribute this heating to eddy currents caused by the magnetic field pulse.
    
    The $H_{c2}$ values were fit to an anisotropic two-band model \cite{Gurevich2003, Kano2009} found to be valid for this material at an almost identical doping (results shown with the solid line), and the measured deviation in temperature has been used to shift the temperature of data points in \cref{fig:results} in the main text.
    
    \begin{figure}
        \centering
        \includegraphics[width=\columnwidth]{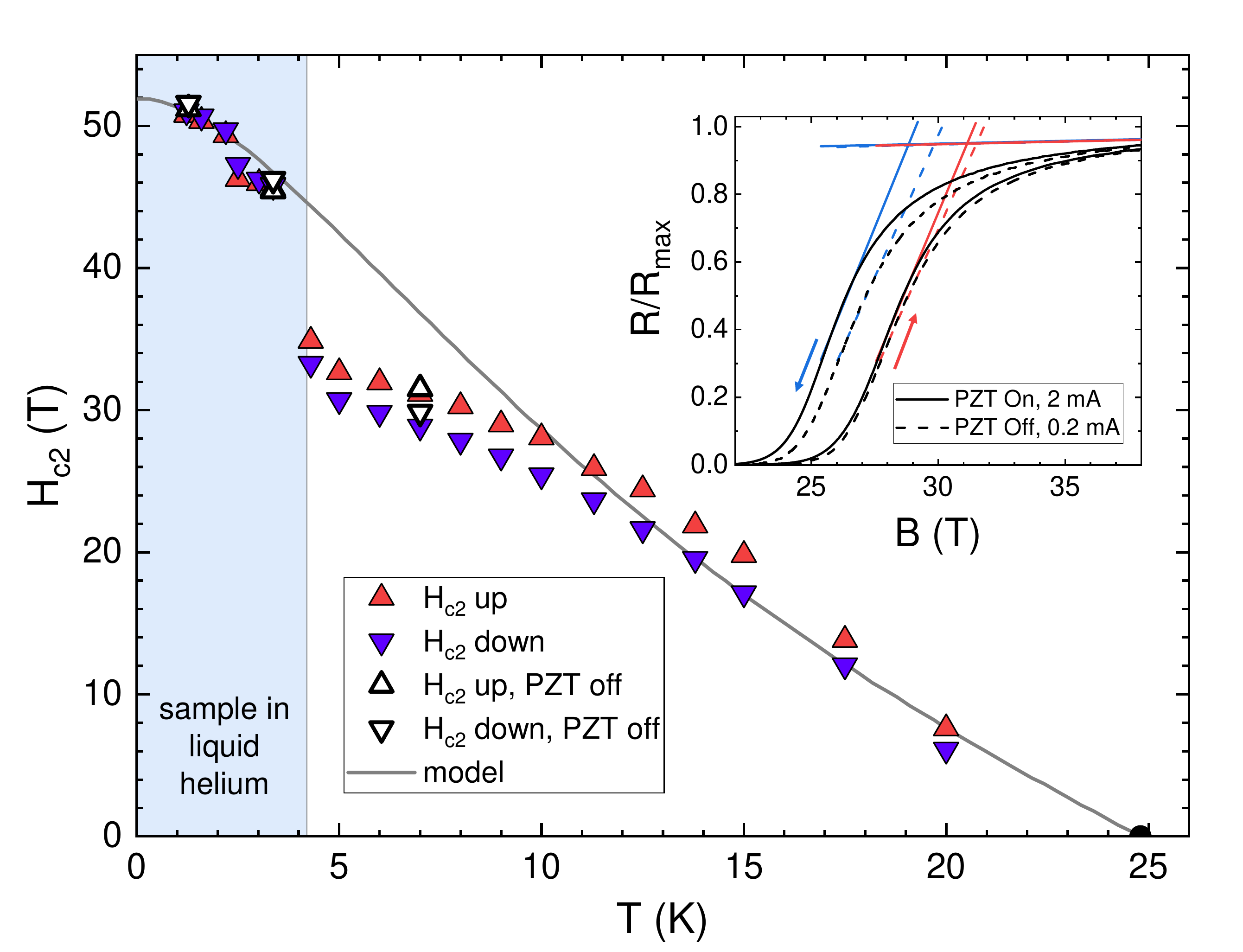}
        \caption{$H_{c2}$ as a function of temperature, measured with increasing field before the strain cycles (upward triangles) and decreasing field after the strain cycles  (downward triangles).  To check for Joule heating within the sample in the resistive phase, several measurements (open symbols) were taken without driving the PZT and with the sample current decreased by a factor of 10.  We observe an increase in $H_{c2}$ when the sample is submerged in liquid helium. 
        Inset: Normalized resistance vs field for two 60~T pulses at 7~K. $H_{c2}$ is extracted from the intersection of the two straight line fits.  Dotted lines corresponds to the data taken without energizing the PZT and using 10\% of the sample excitation current. 
        }
        \label{fig:hc2}
    \end{figure}

\section{Strain accuracy}\label{app:strainxz}
	As described in the main text, the strain measurement during the pulse is performed using a fiber Bragg grating (FBG).  The sample is adhered to the $xy$ plane of the PZT stack, perpendicular to the magnetic field, while the grating runs along the $z$ axis, parallel to the field.  The poling axis of the PZT stack is oriented along the $y$ axis.  In this orientation, the sample (sensitive primarily to strain along the $x$ axis, $\varepsilon_{xx}$) and the FBG (sensitive to $\varepsilon_{zz}$) should both experience the same magnitude and sign of strain for a given voltage applied to the PZT, by merit of both being perpendicular to the poling axis.  In order to verify this assumption, we adhered two resistive strain gauges to the $xy$ and $yz$ planes of another PZT stack from the same manufacturing batch and measured the strain per volt characteristics along both directions.
	
\begin{figure}
		\centering
		\includegraphics[width=\columnwidth]{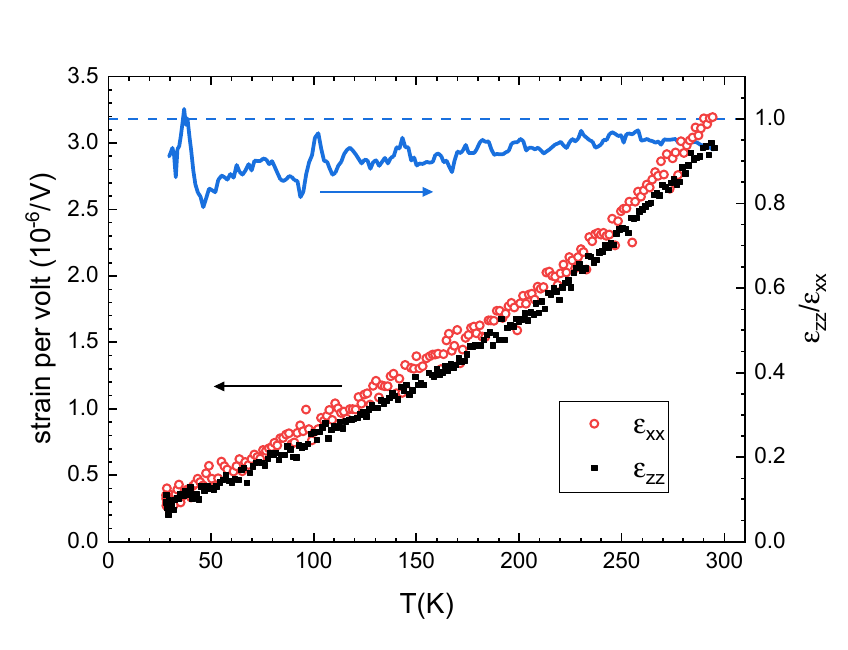}
		\caption{
			Strain per volt measured on a Pst150 2x3x5 PZT stack along both the $x$ and $z$ directions, both perpendicular to the poling axis $y$.
			The sample is primarily sensitive to the $x$ direction, while the FBG detected strain along the $z$ axis, parallel to the field.
			The intrinsic response of the PZT stack itself is expected to be identical in both $x$ and $z$ directions, with an approximately 10\% decrease in the measured $\varepsilon_{zz}$, likely due to the placement of the electrodes along the $yz$ faces of the stack.
		}
		\label{fig:vxz}
	\end{figure}

	The results, in \cref{fig:vxz}, show that the two are indeed very close, with a deviation of approximately 5\%.  The slight suppression of $\varepsilon_{zz}$ relative to $\varepsilon_{xx}$ may be caused by the construction of the stack, which places the electrodes along the $yz$ planes and may stiffen the stack slightly against deformation along $z$.
	
\section{Error Analysis}\label{app:errors}
	The error in the measured gauge factor, $\sigma_{G}$, depends on the measurement error of the oscillating resistivity response to strain amplitude, $\sigma_{\Delta R}$, normalization, $\sigma_{R_0}$, and the oscillating strain amplitude, $\sigma_{\varepsilon}$,
	\begin{equation}
		\left(\frac{\sigma_{G}}{G}\right)^2 = \left(\frac{\sigma_{\Delta R}}{\Delta R}\right)^2 + \left(\frac{\sigma_{R_0}}{R_0}\right)^2 + \left(\frac{\sigma_{\varepsilon}}{\varepsilon}\right)^2
	\end{equation}
	The error in the normalization, $R_0$, is dominated by the magnetoresistance of the sample since the elastoresistivity measurements are performed over a range of fields above $H_{c2}$. $R_0$ is taken to be the average resistance value during the course of the measurement.  The maximum variation during a measurement is 2\% and occurs at the lowest temperatures and highest fields. We use this 2\% value as an upper bound for all temperatures. 

\begin{figure}
        \centering
        \includegraphics[width=8cm]{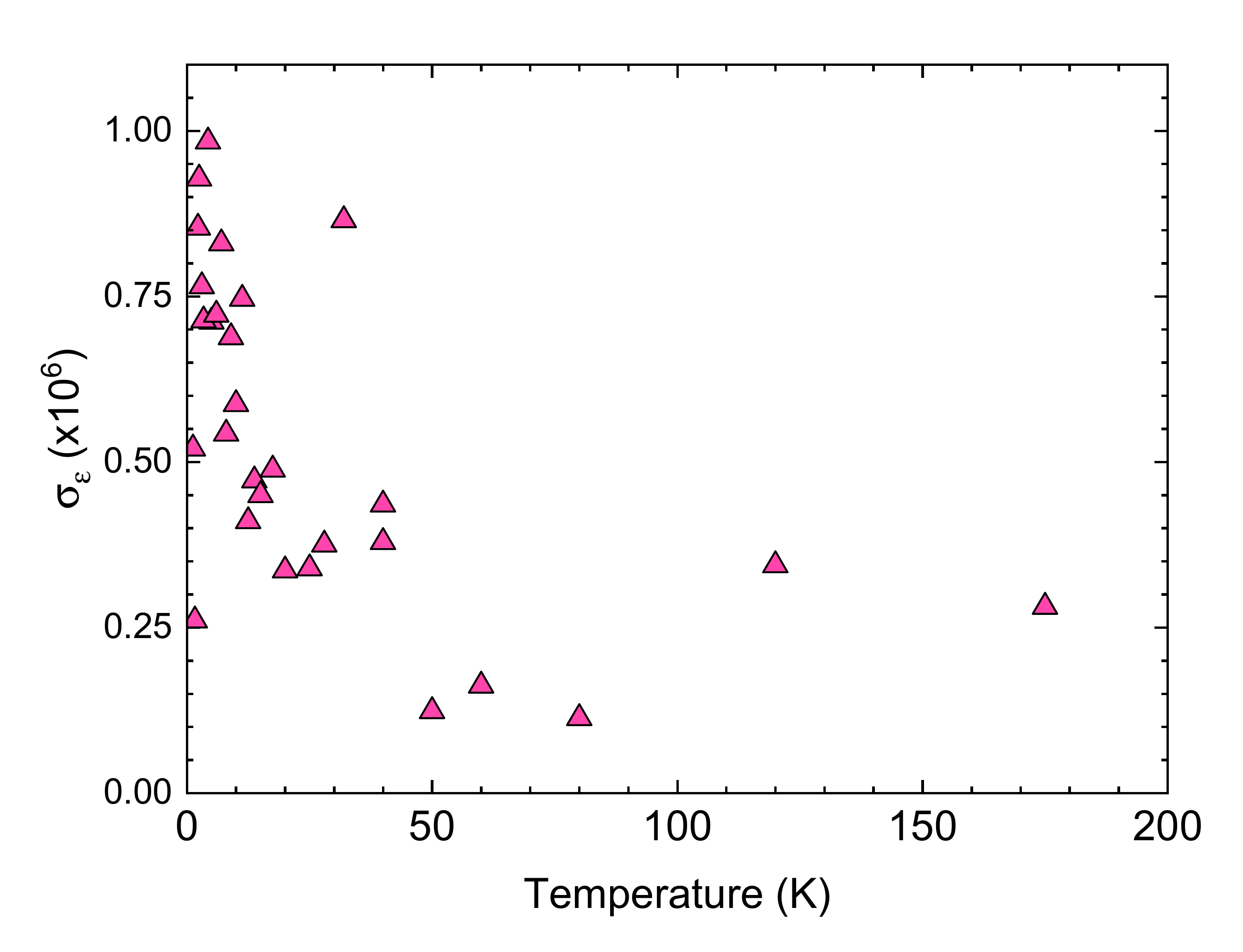}
        \caption{Standard deviation of $\varepsilon_{xx}$ from sets of pulses taken at each temperature.}
        \label{fig:error_strain}
    \end{figure}

\begin{figure}
        \centering
        \includegraphics[width=8cm]{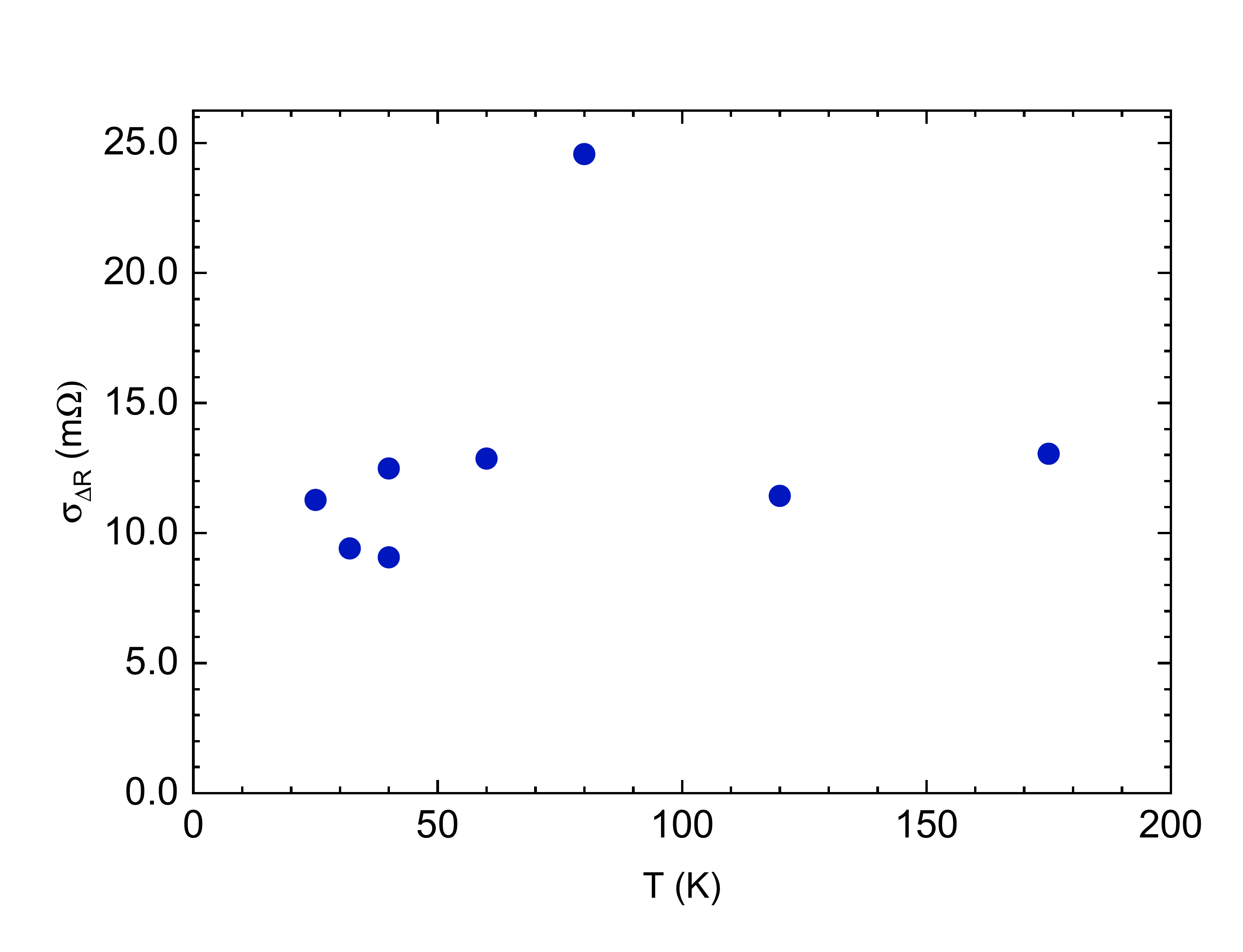}
        \caption{Standard deviation of $\Delta R$ from zero field pulses taken above the superconducting transition. The standard deviation is roughly temperature independent.}
        \label{fig:error_DR}
    \end{figure}

	At every temperature we measured a magnetic field pulse we also performed between 5 and 15 test ``pulses'' in zero field, in which the sample and PZT are driven using the same protocol but the magnet is not fired.  The standard error of the strain for each set of pulses is shown in \cref{fig:error_strain}.  The error increases at colder temperatures. The standard error of the elastoresistivity response can only be measured above the superconducting transition temperature due to practical limitations on the number of high field pulses that can be performed.  The standard error of the modulated resistance amplitude is shown in \cref{fig:error_DR}. It is roughly temperature independent so we use a pooled error of all pulses in the definition of the error bars in \cref{fig:results}.

\section{Comparisons of the temperature dependence of past studies}\label{app:historical}
\begin{figure}
		\centering
		\includegraphics[width=8cm]{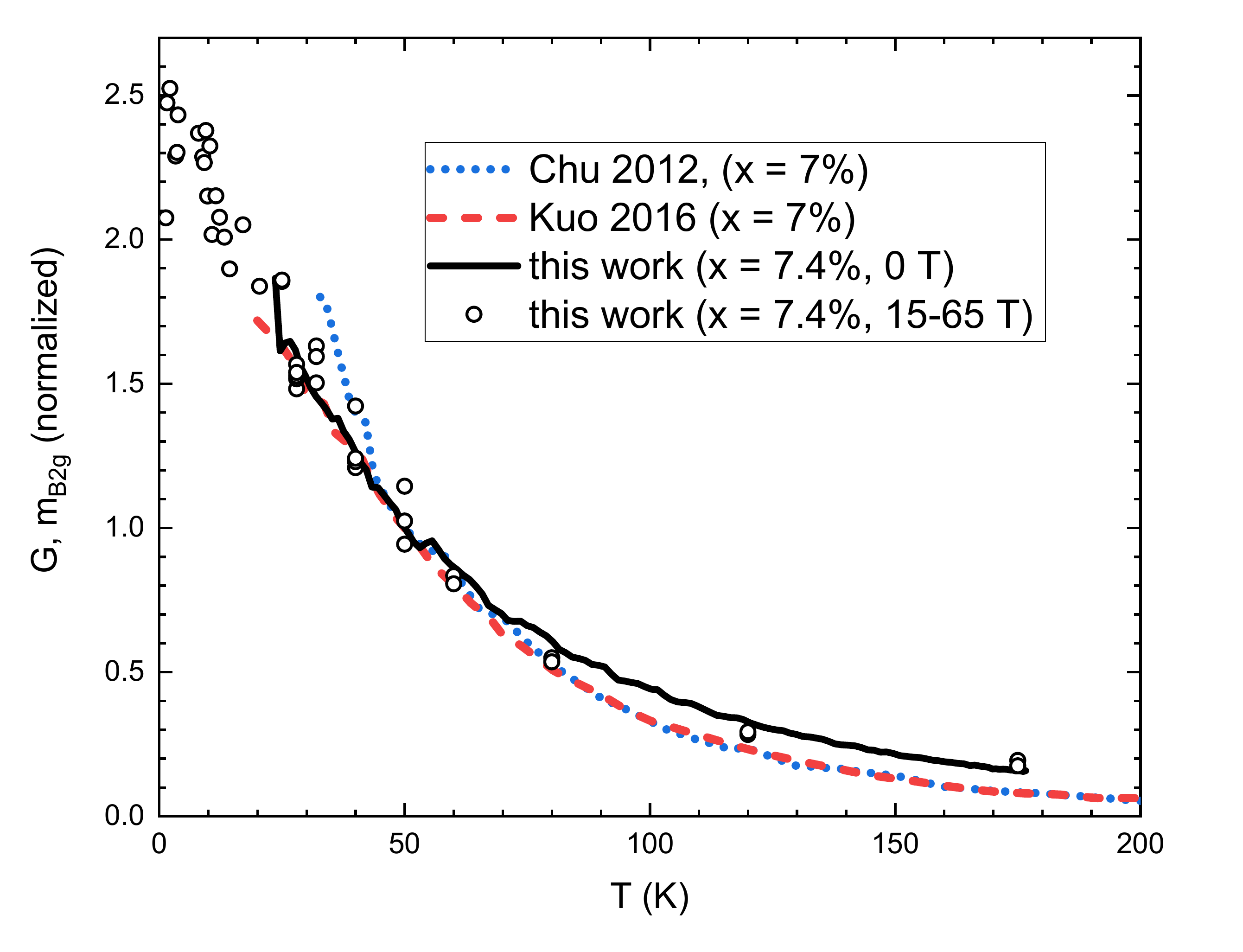}
		\caption{Reported values of $m_{B2g}$ and the gauge factor normalized to the value at 50~K (chosen arbitrarily) for Ba(Fe$_{1-x}$Co$_x$)$_2$As$_2$ from three different studies: Chu 2012\cite{Chu2012}, Kuo 2016\cite{Kuo2016}, and this work.
		Relative to the scale factor used for this work, data from Kuo and Chu require a scaling factor of 1.91 and 0.83, respectively.}
		\label{fig:historical}
	\end{figure}

	Two other published studies to date have presented measurements of the elastoresistivity o	f Ba(Fe$_{1-x}$Co$_x$)$_2$As$_2$ for similar compositions: Chu 2012\cite{Chu2012}, which presents a gauge factor similar to this work, and Kuo 2016\cite{Kuo2016} in which $m_{B_{2g}}$ is measured by extracting $\rho_{xx}$ and $\rho_{yy}$ independently.  To compare all data on the same footing, we normalize all three measurements to their values at 50~K.
	The results, shown in \cref{fig:historical}, show that the temperature dependences of all three measurements find the same qualitative behavior, despite the differences in their implementation.

\section{Testing goodness of fit to Curie-Weiss functional form}\label{app:cw}
	The $B_{2g}$ component of the elastoresistivity tensor for heavily underdoped compositions of Ba(Fe$_{1-x}$Co$_x$)$_2$As$_2$ follows a Curie-Weiss functional form for a region of temperatures above the critical temperature of the structural phase transition \cite{Kuo2016}.  While such behavior  is anticipated for a thermally driven coupled nematic/structural phase transition\cite{Karahasanovic2016}, the behavior of a metal at or near a nematic QCP is essentially unknown \cite{Lederer2015, Schattner2016, Lederer2017, Berg2019}.  Motivated in part by extrapolation from the underdoped compositions, and in part  by the growth of the measured gauge factor at low temperatures, we test here the goodness of fit of the measured data obtained in our pulsed field experiment to the Curie-Weiss functional form. As we show below, while the data plausibly follow Curie-Weiss behavior over a restricted window of intermediate temperatures, the data rise less rapidly than would be expected for a Curie-Weiss behavior at lower temperatures.
	
\begin{figure}
		\centering
		\includegraphics[width=8cm]{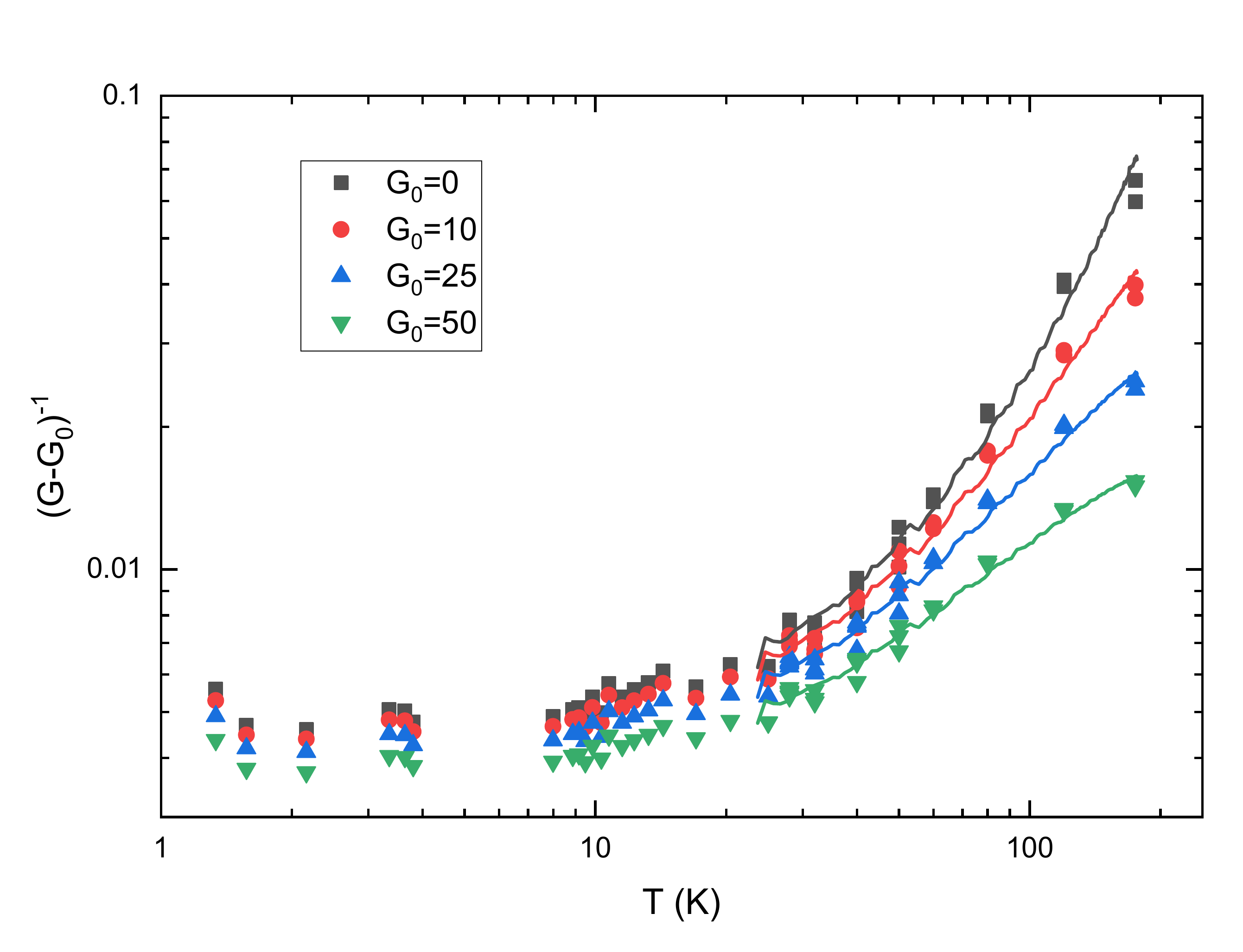}
		\caption{The same data presented in \cref{fig:results}, replotted as its inverse for several values of the temperature independent offset term $G_0$.  If a single set of parameters could describe all of our data with a Curie-Weiss fit, the data would appear linear.  Even the largest $G_0$ presented here, which is unphysically large for Ba(Fe$_{1-x}$Co$_x$)$_2$As$_2$, is unable to reconcile both the high and low temperature behavior with Curie-Weiss scaling.}
		\label{fig:cwfit}
	\end{figure}

	The Curie-Weiss dependence for the nematic susceptibility can be written as
	\begin{equation}
		\chi_N(T) = \frac{C}{T-\Theta} + \chi_0
	\end{equation}
	where $C$ is the Curie constant, $\Theta$ is the Weiss temperature (which would vanish at a QCP), and $\chi_0$ is the susceptibility in the high temperature limit.
	As described in the main text, the gauge factor $G=(-\Delta R/R_0)/\varepsilon_{xx}$ is dominated by a term proportional to the nematic susceptibility, and could be expressed as
	\begin{equation}
		G(T) = \frac{C^*}{T-\Theta} + G_0
	\end{equation}
	The high-temperature limit $G_0$ arises only from geometric effects and should therefore be of order unity for a metal like as Ba(Fe$_{1-x}$Co$_x$)$_2$As$_2$.
	Even with the extension of the available data range by an entire decade at low temperatures presented in this work, any attempt at a three-parameter fit to our data is poorly constrained.
	We illustrate the deviation from this power law scaling at low temperatures by plotting the inverse of the gauge factor data in \cref{fig:cwfit}, in the form
	\begin{equation}
		\frac{1}{G-G_0} = \frac{T-\Theta}{C^*}
	\end{equation}
	for several possible values of $G_0$.
	A quantity which obeys the Curie-Weiss temperature dependence should, on this plot, appear linear over the full range for some value of $G_0$, with slope set by $C^*$ and $y$-intercept set by $\Theta$.
	However, even an unphysically large value of $G_0=50$ cannot reconcile the power law dependence of the inverse gauge factor at high temperature with the relatively flat temperature dependence at low temperatures.
	Such large gauge factors are possible for semiconductors with low carrier densities, where alterations of the band structure and carrier mobilities due to strain can dominate, but are not expected in metals such as Ba(Fe$_{1-x}$Co$_x$)$_2$As$_2$.\cite{Sun2010}
	Only positive values of $G_0$ are presented, as negative values only serve to increase the curvature on this plot.

\end{document}